\documentstyle[12pt]{article}
\def\bbox{{\,\lower0.9pt\vbox{\hrule \hbox{\vrule height 0.2 cm  

\hskip 0.2 cm 

\vrule  height 0.2 cm}\hrule}\,}}
\begin{document}
\setlength{\unitlength}{1mm}
{\hfill  
{\small ALBERTA-THY 17-96, DSF-T-23/96, hep-th/9605153}} \vspace{1cm}\\
\begin{center}
{\Large\bf Cones, Spins and Heat Kernels}
\end{center}
\bigskip\bigskip\bigskip
\begin{center}
{{\bf Dmitri V.~Fursaev}$^{1,2}$ and {\bf Gennaro Miele}$^{3}$}
\end{center}

\bigskip\bigskip

\begin{center}
{\it $^{1}$ Theoretical Physics Institute, Department of Physics,  
University of Alberta,\\ Edmonton, Canada T6G 2J1}
\end{center}
\begin{center}
{\it $^{2}$ Joint Institute for Nuclear Research, Bogoliubov
Laboratory of Theoretical Physics,\\ Dubna Moscow Region, Russia}
\end{center}
\begin{center}
{\it $^{3}$ Dipartimento di Scienze Fisiche, Universit\`a di Napoli - 
Federico II -, and INFN\\ Sezione di Napoli, Mostra D'Oltremare Pad.  
20, 80125, Napoli, Italy}
\end{center}
\vspace*{2cm}
\begin{abstract}
The heat kernels of Laplacians for spin 1/2, 1, 3/2 and 2 fields, and
the asymptotic expansion of their traces are studied on manifolds
with conical singularities. The exact mode-by-mode analysis is carried
out for 2-dimensional domains and then extended to arbitrary
dimensions. The corrections to the first Schwinger-DeWitt coefficients
in the trace expansion, due to conical singularities, are found for
all the above spins. The results for spins 1/2  and 1 resemble the
scalar case. However, the heat kernels of the Lichnerowicz spin 2 operator
and the spin 3/2 Laplacian show a new feature. When the conical angle deficit vanishes
the limiting values of these  traces differ from the
corresponding values computed on the smooth manifold. The reason for
the discrepancy is breaking of the local
translational isometries near a conical singularity. As an application, the results are used to find the ultraviolet divergences in the quantum corrections to the black hole entropy for all these spins.
\end{abstract}
\vspace*{1cm}
\noindent
\begin{center}
{\it PACS number(s): 04.60.+n, 12.25.+e, 97.60.Lf, 11.10.Gh}
\end{center}
\vspace*{2cm}
\noindent
e-mail: dfursaev@phys.ualberta.ca; miele@axpna1.na.infn.it
\newpage
\baselineskip=.6cm

\section {Introduction}
\setcounter{equation}0

Manifolds with conical singularities appear in different physical
applications. The well-known examples are the physics of cosmic strings
\cite{V}, compactifications in the superstring theory \cite{GSW}, and
off-shell computations of the black hole entropy \cite{BTZ}-\cite{MS} addressing its statistical-mechanical origin. Such problems involve
manifolds ${\cal M}_\beta$ where the line element
near the singularity has the form 
\begin{equation}\label{i0}
ds^2=u(r)d\tau ^2+dr^2+\gamma_{ab}(r,y)dy^ady^b~~~.
\end{equation}
Here $\tau$ is a cyclic coordinate $0\leq\tau\leq\beta$, $r\geq 0$ is
a radial coordinate and $u(r)\simeq r^2$ at $r\rightarrow 0$. The
structure of ${\cal M}_\beta$ near the singular hypersurface $\Sigma$,
at $r=0$, is ${\cal C}_\beta\times\Sigma$, where ${\cal C}_\beta$ is a
conical space. We normalize $\tau$ in such a way that at $\beta=2\pi$
the conical singularity is zero and the space is smooth. The important
feature of ${\cal M}_\beta$ is that near $\Sigma$ the components of
the Riemann tensor can be defined only as distributions. The parameter
$\beta$ can be related to the tension of a cosmic string \cite{V}, or,
as in black hole thermodynamics \cite{GiHa:76},\cite{Hawk:79}, it can
be associated with the inverse temperature and in this case $\Sigma$ is the
Euclidean horizon. 

Quantum effects in the presence of conical singularities have been
studied from different points of view and applied to several physical
situations, see for instance \cite{strings1a} - \cite{strings1d}. The
important quantity needed to calculate the effective action in the
theories on curved backgrounds is the trace of the heat kernel
operators $\mbox{Tr}K(s)=\mbox{Tr}\left[e^{-s\bigtriangleup }\right]$.
The properties of $\mbox{Tr}K$ on cones have been studied in detail
for the scalar fields in \cite{Donnelly} - \cite{Dowker94} and
recently for spins 1 and 1/2 in \cite{Kabat},\cite{LW}. It is an
interesting fact that the trace of the heat kernel operators on ${\cal
M}_\beta$ turns out to be a well-defined integral despite the 
fact that integral
characteristics constructed of the powers of the Riemann tensor do not
have in general a strict meaning \cite{FS95b}. 

In this paper we analyze $\mbox{Tr}K$ for Laplacians
$\bigtriangleup^{(j)}$ which appear under quantization of all
physically interesting spins $j$. Our main aim is to find the
modification of the coefficient $A_1^{(j)}$ in the asymptotic
expansion 
\begin{equation}\label{i1}
\mbox{Tr}K^{(j)}(s)=\mbox{Tr}\left[e^{-s\bigtriangleup^{(j)} }\right]  
={1 \over (4\pi s)^{d/2}}\left( A_0^{(j)}+s A_1^{(j)}  +  
s^{2}A_{2}^{(j)}+....\right)~~~
\end{equation}
on the spaces with metric (\ref{i0}) which is important for the
analysis of the ultraviolet divergences and the conformal anomalies in
quantum theory. Spins 2 and 3/2 fields remain an interesting research
subject and we will show that $\mbox{Tr}K^{(j)}$ in these cases have
different properties than the trace for other spins. The point is that the
Laplace operators for spins 2 and 3/2  are sensitive to the isometries
of the background manifold. Locally the manifolds possess
translational symmetries which are broken by conical singularities. It
results in changing the properties of the  heat kernel operators near
the singular hypersurface $\Sigma$ in such a way that even in the
limit $\beta\rightarrow 2\pi$ their traces differ from the
corresponding traces on smooth manifolds. It turns out that in this
limit conical singularities give corrections in the diagonal element
of the heat kernels having the form of a delta-function concentrated
on the hypersurface $\Sigma$. 

In this paper we also discuss the off-shell
computations of the entropy of quantum fields on black-hole
backgrounds. Our new result is the ultraviolet divergent quantum corrections to the entropy for spins 3/2 and 2. We show that for spin 2, 
contrary to other spins,
the properties of the Lichnerowicz operator on singular manifolds 
result to the entropy divergences which cannot be removed under the standard renormalization of the Newton constant.

The paper is organized as follows. In Sec. 2 we define the operators
$\bigtriangleup^{(j)}$, outline the strategy for the computation of
the coefficients $A_1^{(j)}$, and briefly summarize the results.
Sec.'s 3 and 4 are devoted to the explicit derivation of
$A_1^{(j)}$ for spins 1, 2, and 1/2, 3/2, respectively. In particular,
we first analyze the cones ${\cal C}_\beta$ and the two-dimensional
spherical domains $S^2_\beta$ with conical singularities and then make a generalization to arbitrary
dimensions. In Sec. 5 we compare the properties of $A_1^{(j)}$ on
singular spaces ${\cal M}_\beta$ and on the corresponding manifolds
with blunted singularities. Here we also compute the divergent quantum corrections to the entropy for all considered spins on black-hole backgrounds and discuss their renormalization.
The conclusions are presented in Sec. 6. The explicit
computation of the spectrum of the Dirac operator on $S^2_\beta$ is
left for the Appendix. 

\section{Definitions and results}
\setcounter{equation}0

The wave operator $\triangle^{(j)}$ for the fields of different spins
$j$ is defined as follows \cite{CD}. On the spin-$\frac 12$ field
$\psi$ this operator acts as 
\begin{equation}\label{i2}
\bigtriangleup^{(1/2)}\psi=-(\gamma^\alpha\nabla_\alpha)^2\psi=
\left(-\nabla^\alpha\nabla_\alpha+\frac 14R\right)\psi~~~,
\end{equation}
where $\gamma^\mu$ are the Dirac $\gamma$-matrices, and for the vector field $V_\mu$ it reads
\begin{equation}\label{i3}
\bigtriangleup^{(1)} V_\mu=
\left( -\nabla^\alpha\nabla_\alpha \delta_\mu^\nu +R_\mu^\nu  
\right)V_\nu~~~.
\end{equation}
Analogously, for the spin-$\frac 32$ field $\psi_\mu$ we have
\begin{equation}
\label{i4}
\bigtriangleup^{(3/2)}\psi_\mu=-(\gamma^\alpha\nabla_\alpha)^2\psi_\mu 
=\left[\left(-\nabla^\alpha\nabla_\alpha +\frac14 R\right)  
\delta_{\mu}^{\nu}-\frac12 R^\nu_{~\mu\rho\sigma} \gamma^{\sigma}  
\gamma^{\rho}\right]\psi_\nu~~~,
\end{equation}
and for the spin-2 field $h_{\mu\nu}$ 
\begin{equation}
\label{i5}
\triangle^{(2)} h_{\mu\nu}=\left(-\nabla^\alpha\nabla_\alpha  
\delta_{\mu}^ \rho \delta_\nu^ \sigma
+R_\mu^\rho \delta_\nu ^\sigma +R_\nu^\sigma \delta_\mu^\rho
-2R_\mu~^{\rho}~_\nu~^{\sigma}\right)h_{\rho\sigma}~~~.
\end{equation}
The operator $\bigtriangleup^{(1)}$ is the Hodge-deRham operator
acting on $1$-forms and it appears under quantizing in the gauge
$\nabla_\mu  V^\mu=0$. The operator $\bigtriangleup^{(3/2)}$ is the wave operator for the Rarita-Schwinger field \cite{RS} in the
harmonic gauge $\gamma^\mu\psi_\mu=0$ (see Ref. \cite{Das}). Finally,
the operator $\bigtriangleup^{(2)}$ coincides with the Lichnerowicz
operator, which rules the dynamics of small perturbations $h_{\mu\nu}$
of the metric $g_{\mu\nu}$ in the linearized Einstein equations, when
the harmonic gauge $\nabla^\mu(h_{\mu\nu}-\frac 12
g_{\mu\nu}h^\sigma_\sigma)=0$ is fixed \cite{GP}. Note that, all the
above operators can be represented in the form
$\triangle^{(j)}=-\nabla^\alpha\nabla_\alpha+X^{(j)}$, where the
matrices $X^{(j)}$ play the role of potential terms, and are linear in
the curvature of the background space. For this reason in order to have
the well-defined operators we suppose that $X^{(j)}$ are defined in the
regular domain ${\cal M}_\beta-\Sigma$ of the background space (\ref{i0}), and thus they do not include any singular terms.

Let us denote the coordinates $r,\tau, y^a$ in (\ref{i0}) as the {\it
polar coordinates}, thus the tetrades which are parallel to this
coordinate basis will be called {\it polar tetrades} (cf
\cite{Dowker1987}). These tetrades are particularly relevant since
they perform the complete rotation when $\tau$ is increased by $\beta$.
For this reason the connections in the operators $\triangle^{(j)}$
will be computed for polar tetrades, and in this case the
(anti)periodicity conditions for fields $\Phi=(\psi, V_\mu, \psi_\mu,
h_{\mu\nu})$ have the standard form 
\begin{equation}\label{bcond}
\Phi(\tau+\beta,r,y^a)=(-1)^{2j}\Phi(\tau,r,y^a)~~~,
\end{equation}
provided the tensor components are defined in the polar coordinates. 

It is
well-known that on a regular compact manifold ${\cal M}$ the coefficients
$A_1^{(j)}$ in the asymptotic heat kernel expansion (\ref{i1}) can be
written as \cite{CD} 
\begin{equation}\label{i7}
A_1^{(j)}=\int_{\cal M}\left[{N^{(j)} \over 6} R  
-\mbox{Tr}_i\left(X^{(j)}P^{(j)}\right)\right]~~~.
\end{equation}
Hereafter $\mbox{Tr}_i$ denotes the trace over the indices, $P^{(j)}$
is the projector on the corresponding representation of the Lorentz
group and $N^{(j)}=\mbox{Tr}_iP^{(j)}$ is its dimension. If $d$ is the
dimension of space ${\cal M}$, then for the Dirac spinors
$N^{(1/2)}=2^{[d/2]}$, for vector fields $N^{(1)}=d$, and for spin
$3/2$ fields $N^{(3/2)}=2^{[d/2]}d$. In these cases $P^{(j)}$ is the
unit matrix. For rank-2 symmetric tensors $(P^{(2)})^{\rho\sigma}_{\mu\nu}
=\frac 12(\delta^\rho_\mu\delta^\sigma_\nu+\delta^\sigma_\mu\delta^\rho_\nu)$
and $N^{(2)}=d(d+1)/2$. 

On the background manifold ${\cal M}_{\beta}$
with a set $\Sigma$ of singular points, Eq. (\ref{i0}), the
coefficient $A_1^{(0)}$ for the scalar operator $\triangle^{(0)}=-\nabla^\mu\nabla_\mu$ reads
\cite{CKV,F94a,Dowker94} 
\begin{equation}\label{i6}
\bar{A}_1^{(0)}=A_1^{(0)}+A_{\beta,1}^{(0)}~~~,
\end{equation}
\begin{equation}\label{i6a}
A_1^{(0)}=\frac 16 \int_{{\cal M}_\beta-\Sigma} R~~~,~~~A_{\beta,1}^{(0)}=  
{\beta \over 6}
\left(\left ({2\pi \over \beta}\right)^2 - 1\right)\int_{\Sigma}~~~. 
\end{equation}
Here $R$ denotes the Riemann curvature 
calculated in the regular domain ${\cal M}_{\beta}-\Sigma$, and $\int_{\Sigma}$ is the volume of
$\Sigma$ (in two dimensions $\int_{\Sigma}$ is the number of singular
points). 

We will demonstrate that for arbitrary spins the above coefficient on
${\cal M}_\beta$ has an expression similar to the one of the scalar
case (\ref{i6}), and can be represented as the sum 
\begin{equation}\label{i8}
\bar{A}_1^{(j)}=A_1^{(j)}+A_{\beta,1}^{(j)}~~~.
\end{equation}
The term $A_1^{(j)}$ is given by the integral (\ref{i7}) over the
smooth domain of ${\cal M}_\beta$. The presence of the conical
singularities results in the additional term $A_{\beta,1}^{(j)}$,
proportional to the volume of the singular surface $\Sigma$. The precise
form of $A_{\beta,1}^{(j)}$ for the different spins is the following 
\begin{eqnarray}
A_{\beta,1}^{(1/2)}&=&-{N^{(1/2)} \over  
2}A_{\beta,1}^{(0)}~~~,\label{i9}\\
A_{\beta,1}^{(1)}&=&N^{(1)}A_{\beta,1}^{(0)}+2(\beta-2\pi)\int_{\Sigma 
}~~~,\label{i10}\\
A_{\beta,1}^{(3/2)}&=&-{N^{(3/2)} \over 2}A_{\beta,1}^{(0)}+2\beta  
N^{(1/2)}\int_{\Sigma}~~~,\label{i11}\\
A_{\beta,1}^{(2)}&=&N^{(2)}A^{(0)}_{\beta,1}+(2(d+2)
(\beta-2\pi)+8\pi)\int_{\Sigma}~~~,\label{i12}
\end{eqnarray}
where $A_{\beta,1}^{(0)}$ is given by formula (\ref{i6a}). Note that
for spins 2 and 3/2 the surface contributions do not vanish even when
$\beta=2\pi$. 

The reason why $A_{\beta,1}^{(j)}$ have the structure
(\ref{i8}) can be explained as follows. Let $\Sigma_\epsilon$ be a
small domain of thickness $\epsilon$ including $\Sigma$, then
the trace of the heat kernel can be represented
as the sum 
\begin{eqnarray}
\mbox{Tr} K^{(j)}(s)&=&\int_{{\cal M}_\beta}d^dx\sqrt{g}~  
\mbox{Tr}_i K^{(j)}(x,x,s)\nonumber\\
&=&\int_{{\cal M}_\beta-\Sigma_\epsilon}d^dx  
\sqrt{g}~\mbox{Tr}_i  
K^{(j)}(x,x,s)+\int_{\Sigma_\epsilon}d^dx\sqrt{g}~ 
\mbox{Tr}_i K^{(j)}(x,x,s)~~~.\label{D14}  
\end{eqnarray}
The first integral in the r.h.s. of the above equation goes over the
region ${\cal M}_\beta-\Sigma_\epsilon$ which does not include the
conical singularities. Since in this domain one can use for
$\mbox{Tr}_i K^{(j)}(x,x,s)$ the standard Schwinger-DeWitt asymptotic
expansion, the above integral approaches the term $A_1^{(j)}$ given by
Eq. (\ref{i7}) when $\epsilon\rightarrow 0$. On the other hand, in
this limit the second term in the r.h.s. of (\ref{D14}) becomes an
integral over $\Sigma$. From Eq. (\ref{i1}) it follows that
$A_{\beta,1}^{(j)}$ has the dimensionality $L^{d-2}$, where
$L$ is a length (the proper time parameter $s$ in (\ref{i1}) has the
dimensionality $L^2$). The dimensionality of $\int_{\Sigma}$ is $L^{d-2}$, and so
the only form which $A_{\beta,1}^{(j)}$ can have is 
\begin{equation}
A_{\beta,1}^{(j)}=f^{(j)}(\beta)\int_{\Sigma}~~~,\label{aa1}
\end{equation}
where $f^{(j)}(\beta)$ is a dimensionless function which has to be found.  
As a result, $A_{\beta,1}^{(j)}$ cannot depend on the  
curvature of ${\cal M}_\beta$ near $\Sigma$, so in order to determine
its expression the space  
${\cal M}_\beta$ can be safely approximated  by ${\cal C}_\beta\times\Sigma$
\begin{equation}\label{i14}
\left.A_{\beta,1}^{(j)}\right|_{{\cal M}_\beta}=\left.A_{\beta,1}^{(j)}\right|_{{\cal  
C}_\beta\times\Sigma}~~~.
\end{equation}
To compute $A_{\beta,1}^{(j)}$ on spaces where $\partial/
\partial \tau$ is the globally defined Killing field we will also  
use the formula
\begin{equation}\label{i15}
A_{\beta,1}^{(j)}=\left.\bar{A}_1^{(j)}\right|_{{\cal M}_\beta}-{\beta \over  
2\pi}\left.
A_1^{(j)}\right|_{{\cal M}_{\beta=2\pi}}~~~.
\end{equation}
The last term in (\ref{i15}) subtracts the contribution of the smooth
domain ${\cal M}_\beta-\Sigma$. These contributions for  ${\cal M}_\beta$ and for the 
corresponding manifold ${\cal
M}_{\beta=2\pi}$ without singularities differ from each other by the coefficient $\beta/2\pi$,
which is related to the different periods in $\tau$ on these spaces. 

\section{Integer spin fields}
\setcounter{equation}0
\subsection{Vector field}

Let us consider the wave operator (\ref{i3}) for the vector field. The
eigen-value problem for this operator has a simple solution in two
dimensions since its eigen-functions can be expressed in terms of the
scalar ones corresponding to the operator $\triangle^{(0)}$. Indeed,
in this case one can use the Hodge-deRham decomposition 
of the vector $V_\mu$ into transverse $V_\mu^{T}$ , longitudinal
$V_\mu^{L}$  and harmonic $V_\mu^H$  parts
\begin{equation}\label{HDdec}
V_\mu=V_\mu^{T}+V_\mu^{L}+V_\mu^H~~~,
\end{equation}
and in two dimension it results
\begin{equation}\label{HDdec2}
V_\mu^{T}=\epsilon_{\mu\nu}\phi^{,\nu}~~~,~~~V_\mu^{L}=\rho_{,\mu}~~~, 
~~~\nabla^\mu V_\mu^H=\epsilon^{\mu\nu}\nabla_\mu V_\nu^H=0~~~,
\end{equation}
where $\epsilon_{\mu\nu} $ is the rank 2 antisymmetric tensor, and $\phi$,  
$\rho$ are two scalar fields. By using (\ref{HDdec2}) one can easily check that 
\begin{equation}\label{11}
\triangle^{(1)}(\rho_{,\mu})=(  
\triangle^{(0)}\rho)_{,\mu}~~~,~~~
\triangle^{(1)}(\epsilon_{\mu\nu}\phi^{,\nu})=\epsilon_{\mu\nu}
( \triangle^{(0)}\phi)^{,\nu}~~~.
\end{equation}
Therefore, from Eq.(\ref{11}) and the decomposition (\ref{HDdec})  
one gets the following relation between the traces
\begin{equation}\label{12}
\mbox{Tr}K^{(1)}=
2~\mbox{Tr}K^{(0)}+n_1-2n_0~~~,
\end{equation}
where $n_1$, $n_0$ are the numbers of zero modes of the operators  
$\triangle^{(1)}$ and $\triangle^{(0)}$ respectively. Note that the
harmonic vectors $V_\mu^H$ are zero modes of $\triangle^{(1)}$.
Equation (\ref{12}) means that all Schwinger-DeWitt coefficients in
the expansion of the vector heat kernel are twice the scalar ones.
The exception is the $A_1$-coefficient which according to  (\ref{12})
obeys a simple index theorem 
\begin{equation}\label{13}
2A_1^{(0)}-A_1^{(1)}=4\pi(2 n_0-n_1)~~~.
\end{equation}
The last result can be represented in another form by making use of  
the following identity
\begin{equation}\label{16}
n_0-n_1+n_2= \chi[{\cal M}]~~~,
\end{equation}
which expresses the Euler characteristics $\chi[{\cal M}]$ of the
background manifold ${\cal M}$ in terms of Betti numbers $n_p$ or,
which is the same, numbers of the harmonic $p$-forms. For a compact
manifold $n_2=n_0$, and Eq. (\ref{13}) reads 
\begin{equation}\label{17}
2A_1^{(0)}-A_1^{(1)}=4\pi\chi[{\cal M}]~~~.
\end{equation}
The expression (\ref{13}) can be strictly proved for smooth  
manifolds. 

In order to get the coefficient $A_1^{(1)}$
for the singular space ${\cal M}_\beta$ we can use the results of  
Ref. \cite{FS95b}, where it has been shown that the Euler characteristics  
on manifolds with conical singularities have well defined expressions.  
In particular, in 2 dimensions
the Euler number can be written in the form \cite{BTZ},\cite{FS95b}
\begin{equation}\label{14}
\chi[{\cal M}_\beta]={1 \over 4\pi} \left( 2(2\pi  
-\beta)\int_{\Sigma}+\int_{{\cal M}_\beta} R \right)~~~,
\end{equation}
(if the manifold has a boundary one must also add a boundary term).  
Thus, by using (\ref{13}) and (\ref{14}) one gets 
\begin{eqnarray}
\bar{A}_1^{(1)}&=&2\bar{A}_1^{(0)}-4\pi\chi[{\cal M}_\beta]=
2\bar{A}_1^{(0)}- 2(2\pi -\beta)\int_{\Sigma}-\int_{{\cal M}_\beta-\Sigma} R
\nonumber\\
&=&A_1^{(1)}+2\left.A_{\beta,1}^{(0)}
\right|_{{\cal C}_\beta}+2(\beta-2\pi)~~~.
\label{15} 
\end{eqnarray}
Consequently, in 2 dimensions the contribution to the vector  
coefficient $\bar{A}_1^{(1)}$ due to the conical singularity is
\begin{equation}\label{16a}
\left.A_{\beta,1}^{(1)}
\right|_{{\cal C}_\beta}=2\left.A_{\beta,1}^{(0)}
\right|_{{\cal C}_\beta}+2(\beta-2\pi)~~~,
\end{equation}
and it agrees with the result of Kabat \cite{Kabat} found by a  
different method. To extend this result to arbitrary dimension we use  
Eq.(\ref{i14}) and calculate this coefficient on the space  
${\cal C}_\beta\times\Sigma$. In this case, one needs first to 
decompose the vector field onto the parts orthogonal and tangent to  
$\Sigma$. According to this decomposition one has
\begin{equation}\label{18}
\left. \mbox{Tr} K^{(1)}\right|_{{\cal C}_\beta\times\Sigma}=
\mbox{Tr}\left.  K^{(1)}\right|_{{\cal C}_\beta}
\left.\mbox{Tr} K^{(0)}\right|_{\Sigma}
+ \mbox{Tr}\left. K^{(0)}\right|_{{\cal C}_\beta}
\left.\mbox{Tr} K^{(1)} \right|_{\Sigma}~~~.
\end{equation}
By taking into account  (\ref{18}) we come to the formula 
\begin{eqnarray}
A_{\beta, 1}^{(1)}& = &\left. A_{\beta,1}^{(1)}\right|_{{\cal  
C}_\beta}\int_{\Sigma}+\left.A_{\beta,1}^{(0)}\right|_ 
{{\cal C}_\beta}(d-2)\int_{\Sigma}\nonumber\\
&=&\left(\left. d A_{\beta,  
1}^{(0)}\right|_{{\cal C}_\beta}+2(\beta-2\pi)\right)\int_{\Sigma}=
 d A_{\beta, 1}^{(0)}+2(\beta-2\pi)\int_{\Sigma}~~~.
\label{21} 
\end{eqnarray}
Eq. (\ref{21}) gives the contribution (\ref{i10}) due to the conical  
singularities to the heat coefficient of the vector operator in
$d$ dimensions.  

\subsection{Lichnerowicz operator}
The Lichnerowicz operator $\triangle^{(2)}$ acting on the symmetric
second-rank tensors $h_{\mu\nu}$ can be obtained by expanding the
Ricci tensor with respect to the perturbation $h_{\mu\nu}$
of the  background metric $g_{\mu\nu}$ \cite{GP}
\begin{equation}\label{L1}
R_{\mu\nu}(g+h)-R_{\mu\nu}(g)=\frac 12 \triangle^{(2)}h_{\mu\nu}+
O(h^2)~~~,
\end{equation}
when the gauge condition $\nabla^\mu(h_{\mu\nu}-\frac 12
g_{\mu\nu}h^\sigma_\sigma)=0$ is imposed. As it can be easily shown,
the quantization of the gravitational field in this gauge leads to the
computation of the determinant of $\triangle^{(2)}$ \cite{CD}. For the
following applications it is convenient to report two properties of
$\triangle^{(2)}$ 
\begin{eqnarray}
\triangle^{(2)}g_{\mu\nu}\phi&=&g_{\mu\nu}\triangle^{(0)}\phi~~~,\label{L2}\\ 
\triangle^{(2)}(\nabla_\mu V_\nu +\nabla_\nu V_\mu)&=&\nabla_\mu
\triangle^{(1)}V_\nu +\nabla_\nu \triangle^{(1)} V_\mu~~~.\label{L3} 
\end{eqnarray}
Eq. (\ref{L2}) holds in general, while (\ref{L3}) is valid
on Einstein spaces where $R_{\mu\nu}=g_{\mu\nu}\Lambda$,  with  
$\Lambda$  denoting a (cosmological) constant. 

In analogy to the spin 1 case, we start our considerations in 2
dimensions. According to (\ref{L3}) the properties of
$\triangle^{(2)}$ are simplified on the constant curvature spaces. For
this reason we consider the compact space $S^2_\beta$ with the metric 
\begin{equation}\label{L4}
ds^2=\cos^2\theta d\tau^2 +d\theta^2~~~,
\end{equation}
where $0\leq\tau\leq\beta$ and $-\pi/2 \leq \theta \leq +\pi/2$. 
Eq. (\ref{L4}) describes a two-dimensional unit sphere with two conical singularities at the poles
$x_1$ and $x_2$ $(\theta=\pm\pi/2)$. It is convenient to study first
this simplest case because $\triangle^{(2)}$ has a compact spectrum on
$S^2_\beta$ which can be found exactly. To this aim, let us remind
that the second rank tensor in two dimensions can be decomposed as
follows \cite{GP} 
\begin{equation}\label{L5}
h_{\mu\nu}=h_{\mu\nu}^{TT}+h_{\mu\nu}^{L}+\frac 12 g_{\mu\nu}  
h^\sigma_\sigma~~~,
\end{equation}
where $h_{\mu\nu}^{TT}$ and $h_{\mu\nu}^{L}$ are the traceless tensors 
\begin{equation}\label{L6}
\nabla^\mu h_{\mu\nu}^{TT}=0~~~,
\end{equation}
\begin{equation}\label{L7}
h_{\mu\nu}^{L}=\nabla_\mu V_\nu +\nabla_\nu  
V_\mu-g_{\mu\nu}\nabla^{\sigma}V_{\sigma}~~~,
\end{equation}
and $V_\mu$ is a vector. We define $\triangle^{(2)}$ on the
second-rank symmetric tensors $h_{\mu \nu}$ obeying the condition
(\ref{bcond}) and having the finite norm 
\begin{equation}\label{L71}
||h||^2=\int_{S^2_\beta}\sqrt{g}d^2x  
~h^{*}_{\mu\nu}h^{\mu\nu}~<~\infty~~~.
\end{equation}
One can show that there are no transverse tensors  
$h_{\mu\nu}^{TT}$ on $S^2_\beta$ obeying these conditions. So by  
taking into account Eqs.(\ref{L2}),(\ref{L3}) and (\ref{L5}) one
can represent the trace of the Lichnerowicz operator as
\begin{equation}\label{L8}
\mbox{Tr} K_\beta^{(2)}(s)=\mbox{Tr} K_\beta^{(0)}(s)+\mbox{Tr}  
K_\beta^{(1)}(s)-
\sum_{l=1}^{n_{ck}}e^{-s\lambda_l}~~~.
\end{equation}
For the reasons which will be clear later, we will write the 
subscript $\beta$ for the heat kernel on $S^2_\beta$ ($\beta\neq  
2\pi$). The last term in the r.h.s. of (\ref{L8}) subtracts from 
the
vector heat kernel $\mbox{Tr} K^{(1)}(s)$ the contribution of 
$n_{ck}$ vector modes, for which the tensor modes $h_{\mu\nu}^{L}$
are identically zero. From the definition (\ref{L7}) it follows that  
such vector modes are the solutions of the two dimensional 
conformal Killing equation
\begin{eqnarray}\label{L9}
\nabla_\mu V_\nu +\nabla_\nu  
V_\mu-g_{\mu\nu}\nabla^{\sigma}V_{\sigma}=0~~~.
\end{eqnarray}
The results of the previous section enable one to rewrite (\ref{L8})  
in the form
\begin{equation}\label{L10}
\mbox{Tr} K_\beta^{(2)}(s)=3\mbox{Tr}  
K_\beta^{(0)}(s)-\chi[S^2_\beta]-\sum_{l=1}^{n_{ck}}e^{-s\lambda_l}~~~ 
,\end{equation}
where $\chi[S^2_\beta]=2$ is the Euler number,
and to get the expression for the first coefficient in the  
asymptotic expansion for $\mbox{Tr} K_\beta^{(2)}(s)$
\begin{equation}\label{L11}
\bar{A}_1^{(2)}=3\bar{A}_1^{(0)}-4\pi(\chi[S^2_\beta]+n_{ck})~~~.
\end{equation}
It is worth examining how this formula works on $S^2$. In this case  
there are 6 solutions of (\ref{L9}): 3 {\it true} Killing fields
$(V^l)_{\mu}=\epsilon_{\mu\nu}\nabla^{\mu}\phi^l$ 
(corresponding to the $SO(3)$ isometry of $S^2$)
and 3 {\it conformal} Killing vectors $({\bar  
V}^l)_\mu=\epsilon_{\mu\nu}(V^l)^{\nu}$. It is easy to check that 
$\phi^l$ are 3 spherical (dipole) eigen-functions $  
\phi^0,\phi^{\pm}$ of the scalar Laplacian on $S^2$, 
\begin{equation}\label{L12}
\triangle^{(0)}\phi^l=2\phi^l~~~,
\end{equation}
\begin{equation}\label{L13}
\phi^0=\sin\theta~~~,~~~\phi^{\pm}=\cos\theta e^{\pm i\tau}~~~.
\end{equation}
Thus, the formula (\ref{L11}) gives for $A_1$-coefficient the  
expression
\begin{equation}\label{L14}
A_1^{(2)}=3A_1^{(0)}-4\pi(2+6)~~~,
\end{equation}
which exactly coincides with general expression (\ref{i7}). Note that
in 2d case $\mbox{Tr}_i(PX^{(2)})=4R=8$. 

Let us consider now the
singular space $S^2_\beta$ with $\beta\neq 2\pi$. In this case, the
conical singularities near the poles break the symmetry $SO(3) \times
SO(3)$, corresponding to the conformal Killing fields on $S^2$, to
$O(1)\times O(1)$ \footnote{The symmetry is unbroken only when
$\beta=2\pi k$, where $k\geq 2$ is a natural number, see a comment in the end of
this section.}. Hence, the number of solutions of (\ref{L9}) reduces
to 2 vectors determined by the scalar mode $\phi^0$, Eq. (\ref{L13}).
The later, as before, is the eigen-mode of $\triangle^{(0)}$. The
other scalar modes $\phi^{\pm}$  transform on $S^2_\beta$ into the functions
$\phi^{\pm}_\beta$ 
\begin{equation}\label{L15}
\triangle^{(0)}\phi_\beta^{\pm}=\alpha(\alpha+1)\phi_\beta^{\pm}~~~,~~ 
~\alpha={2\pi \over \beta}~~~,
\end{equation}
\begin{equation}\label{L16}
\phi^{\pm}_\beta=\cos^{\alpha}\theta e^{\pm i\alpha\tau}~~~.
\end{equation}
The corresponding vector modes
$(V^{\pm}_\beta)_\mu=\epsilon_{\mu\nu}\nabla^{\nu}\phi_\beta^{\pm}$
and $(\bar{V}^{\pm}_\beta)_\mu=\epsilon_{\mu\nu}(V^{\pm}_\beta)^\mu$
are no more the solutions of (\ref{L9}), and, according to (\ref{L7}), one
can construct in terms of them the following second rank tensors 
\begin{eqnarray}
(h^{\pm}_\beta)_{\mu\nu}&=&\frac 12 C(\beta) \left(\nabla_\mu  
(V^{\pm}_\beta)_\nu +\nabla_\nu  
(V^{\pm}_\beta)_\mu\right)~~~,\label{L17}\\
(\bar{h}^{\pm}_\beta)_{\mu\nu}&=&C(\beta)\left(\nabla_\mu  
(\bar{V}^{\pm}_\beta)_\nu +\nabla_\nu  
(\bar{V}^{\pm}_\beta)_\mu-g_{\mu\nu}\nabla^{\sigma}
(\bar{V}^{\pm}_\beta)_\sigma\right)~~~,\label{L18} 
\end{eqnarray}
where $C(\beta)$ is a normalization constant. 
The tensor modes (\ref{L17}) and (\ref{L18}) have the following  
properties: 
\begin{itemize}
\item[i)] they obey the periodicity condition (\ref{bcond}), 
\item[ii)] are the eigen-functions of $\triangle^{(2)}$, 
\item[iii)]  can be normalized on $S^2_\beta$ if $\beta<2\pi$  
($\alpha>1$)
\begin{equation}\label{L19}
||h_\beta^{\pm}||^2=||\bar{h}_\beta^{\pm}||^2=
4\pi^{3/2}(\alpha-1){\Gamma(\alpha+3) \over \Gamma(\alpha+3/2)}
|C(\beta)|^2~~~.
\end{equation}
\end{itemize}
Consequently, at $\beta<2\pi$ the Lichnerowicz operator acquires 4 new
additional eigen-modes $(h^{\pm}_\beta)_{\mu\nu}$ and
$(\bar{h}^{\pm}_\beta)_{\mu\nu}$, which we will call for simplicity the
dipole modes. We will consider first the case $\beta <2\pi$, then
the results can be generalized on arbitrary
$\beta$'s with the help of analytical continuation. 

Obviously, in the presence of the dipole modes the limiting
value of the trace $\mbox{Tr}K^{(2)}_\beta$ when $\beta\rightarrow
2\pi$ does not coincide with the trace $\mbox{Tr}K^{(2)}$ on $S^2$.
Let us discuss this question in more detail. The heat kernel
$K^{(2)}_\beta$ at $\beta<2\pi$ can be represented in the form (for
simplicity we do not write here the tensor indexes) 
\begin{eqnarray}
K^{(2)}_\beta(s)(x,x',s)&=&\tilde{K}^{(2)}_\beta(x,x',s)+D^{(2)}_{\beta} 
(x,x',s)~~~,\label{L20}\\ 
\tilde{K}^{(2)}_\beta(x,x',s)&=&\sum_{\lambda\neq \alpha(\alpha+1)}
h^\lambda(x)^{*}h^\lambda(x')e^{-s\lambda}~~~,\label{L21}\\
D^{(2)}_{\beta}(x,x',s)&=&\sum_{\pm}\left((h^{\pm}_\beta)^{*}(x)h^{\pm}_ 
\beta(x')+(\bar{h}^{\pm}_\beta)^{*}(x)\bar{h}^{\pm}_\beta(x')
\right)e^{-s\alpha(\alpha+1)}\nonumber\\
&\equiv&\sum_{\pm}(f_\beta^{\pm}(x,x')+\bar{f}_\beta^{\pm}(x,x'))
e^{-s\alpha(\alpha+1)}~~~.\label{L22} 
\end{eqnarray}
The quantity $\tilde{K}^{(2)}_\beta$ denotes the part of
$K^{(2)}_\beta$ which does  not include the dipole modes
$h^{\pm}_\beta$, $\bar{h}^{\pm}_\beta$. There is a one-to one
correspondence between the modes in $\tilde{K}^{(2)}_\beta$ and the modes of
the heat-kernel operator $K^{(2)}$ on sphere $S^2$. Therefore, in the
limit $\beta\rightarrow 2\pi$ the both kernels coincide 
\begin{equation}\label{L23}
\lim_{\beta\rightarrow 2\pi}\tilde{K}^{(2)}_\beta(x,x',s)=K^{(2)}
(x,x',s)~~~.\end{equation}
Let us investigate now the same limit for $D^{(2)}_\beta$, defined in  
Eq. (\ref{L22}). The normalization condition for the dipole modes is
\begin{equation}\label{L24}
\int\mbox{Tr}_i f_\beta^{\pm}(x,x)=\int\mbox{Tr}_i  
\bar{f}_\beta^{\pm}(x,x)=
||h_\beta^{\pm}||^2=||\bar{h}_\beta^{\pm}||^2=1~~~.
\end{equation}
Thus, as it follows from (\ref{L19}), the normalization constant
$C(\beta)$ diverges as $(2\pi-\beta)^{-1/2}$ as $\beta\rightarrow
2\pi$. It is easy to see, using Eqs. (\ref{L16}), (\ref{L17}) and
(\ref{L18}), that when $\beta\rightarrow 2\pi$ the functions
$f_\beta^{\pm}$ and $\bar{f}_\beta^{\pm}$ vanish as fast as
$(2\pi-\beta)$ in the all points of $S^2_\beta$ except its two poles
$x_1$ and $x_2$ (at $\theta=\pm \pi/2$). However, near the poles 
$(\det g_{\mu\nu}(x))^{1/2}\mbox{Tr}_if_\beta^{\pm}(x,x)\sim \cos^{2\alpha-3}\theta$ (the
same is true for $\bar{f}_\beta^{\pm}$), and there is a singularity
which is not integrable at $\beta=2\pi$ ($\alpha=1$). The above properties
demonstrate that functions $\mbox{Tr}_if_\beta^{\pm}(x,x)$ are
the
distributions when $\beta=2\pi$ 
\begin{equation}\label{L25}
\lim_{\beta\rightarrow 2\pi}\mbox{Tr}_if_\beta^{\pm}(x,x)=
\lim_{\beta\rightarrow 2\pi}\mbox{Tr}_i\bar{f}_\beta^{\pm}(x,x)=
\frac 12\left(\delta(x,x_1)+\delta(x,x_2)\right)~~~,
\end{equation}
where $\delta(x,x_i)$ are the covariant $\delta$-functions on $S^2$. It
means that the dipole modes in the limit $\beta=2\pi$ give to the heat
kernel a contribution concentrated on the poles. Hence, by making use
of (\ref{L23}) and (\ref{L25}) one can find the following limit  
\begin{equation}\label{L26}
\lim_{\beta\rightarrow 2\pi}\mbox{Tr}_i  
K^{(2)}_\beta(x,x,s)=\mbox{Tr}_i  
K^{(2)}(x,x,s)+2\left(\delta(x,x_1)+\delta(x,x_2)\right)e^{-2s}~~~.
\end{equation}
The factor 2 in the last term in of r.h.s. of Eq. (\ref{L26})
corresponds to the number of the Killing generators which are broken in
the presence of conical singularities. Note that the singular term in
(\ref{L26}) can only appear in integral quantities. Thus, in the
physical applications the last term in (\ref{L26}) will not contribute
to the local observables (for instance, the averages of the field
operators) calculated with the help of the heat kernel
$K^{(2)}_\beta(x,x',s)$. 

Consider now how the conical singularities change the 
$A_1$-coefficient for the Lichnerowicz  
operator. It follows from (\ref{L10}) that 
\begin{equation}\label{L27}
\mbox{Tr} K_\beta^{(2)}(s)=3\mbox{Tr}  
K_\beta^{(0)}(s)-\chi[S^2_\beta]-2e^{-2s}~~~,
\end{equation}
and the complete coefficient looks as
\begin{equation}\label{L28}
\bar{A}_1^{(2)}=3\bar{A}_1^{(0)}-4\pi\chi[S^2_\beta]-8\pi~~~.
\end{equation}
Let $\left.A_{\beta,1}^{(2)}\right|_{{\cal C}_\beta}$ be the
contribution into
$\bar{A}_1^{(2)}$ from one conical singularity.
Two singular points of $S^2_\beta$ give the correction which can be found by making use of (\ref{i15})
\begin{equation}\label{L29}
2\left.A_{\beta,1}^{(2)}\right|_{{\cal C}_\beta}=\bar{A}_1^{(2)}-{\beta \over 2\pi}A_1^{(2)}~~~,
\end{equation}
where $A_1^{(2)}$ is the value of the heat coefficient on $S^2$, Eq.
(\ref{L14}). Then Eqs. (\ref{L14}) and (\ref{L29}) result in the
expression
\begin{equation}\label{L30}
\left.  
A_{\beta,1}^{(2)}\right|_{{\cal C}_\beta}=
3\left.  
A_{\beta,1}^{(0)}\right|_{{\cal C}_\beta}
+8(\beta-2\pi)+ 2\cdot4\pi~~~.
\end{equation}
The dipole modes give in (\ref{L30}) the additional term $2\cdot4\pi$
which survives even in the limit $\beta=2\pi$. This term depends only
on the conical geometry ${\cal C}_\beta$ near a singular point, and
for this reason it must be universal in $A_1$ of
$\mbox{Tr}K_\beta^{(2)}$ for all manifolds with the given kind of
singularities. The factor 2 is the number of Killing generators
(corresponding to translation symmetry) which are broken when the
plane $R^2$ is changed by the cone ${\cal C}_\beta$. Note that the
same number of symmetries are broken in the transition from $S^2$ to
$S^2_\beta$. 

The generalization of (\ref{L30}) to arbitrary
dimensional manifolds is analogous to the vector case. It is
sufficient to consider the space product ${\cal C}_\beta \times
\Sigma$ and to use the relation 
\begin{equation}
\left. \mbox{Tr}~K^{(2)}\right|_{{\cal C}_\beta \times \Sigma}=
\left. \mbox{Tr}~K^{(0)}\right|_{{\cal C}_\beta}
\left. \mbox{Tr}~K^{(2)}\right|_{\Sigma}+
\left. \mbox{Tr}~K^{(2)}\right|_{{\cal C}_\beta} 
\left. \mbox{Tr}~K^{(0)}\right|_{\Sigma}+
\left. \mbox{Tr}~K^{(1)}\right|_{{\cal C}_\beta}
\left. \mbox{Tr}~K^{(1)}\right|_{\Sigma}.
\label{L31}
\end{equation}
This formula follows from the definition (\ref{i5}) of the Lichnerowicz operator and the decomposition of a rank $2$ tensor onto tensors with the components either tangent or orthogonal to $\Sigma$,
and a tensor with the mixed components. By making use of
(\ref{L31}) and the results for the vector field we get 
\begin{equation}
\bar{A}^{(2)}_{\beta,1}=\left. A^{(0)}_{\beta,1}\right|_{{\cal  
C}_\beta}N^{(2)}(d-2)
\int_{\Sigma}+
\left. A^{(2)}_{\beta,1}\right|_{{\cal C}_\beta}
\int_{\Sigma}
 + \left. A^{(1)}_{\beta,1}\right|_{{\cal C}_\beta}(d-2)
\int_{\Sigma}~~~.
\label{L32}
\end{equation}
Thus, by observing that
\begin{equation}
N^{(2)}(d-2) + 3 +2(d-2)=N^{(2)}(d)~~~,
\label{3.2-19} 
\end{equation}
we finally get the formula
\begin{equation}
\bar{A}^{(2)}_{\beta,1}=N^{(2)}\bar{A}^{(0)}_{\beta,1}+(2(d+2)
(\beta-2\pi)+8\pi)\int_{\Sigma}~~~,
\label{3.2-20} 
\end{equation}
already reported in Sec. 2.

On manifolds with the periodicity $\beta=2\pi k$, where  $k$ is a natural 
number $\geq 2$, the Killing vectors are the same as on the
corresponding smooth spaces ($\beta=2\pi$). The properties of
$K^{(2)}$ when $\beta$ approaches $2\pi k$ are also the same and the
investigation of this limit is similar to
the analysis given in this section. 

\section{Fields with half odd-integer spins}
\setcounter{equation}0
\subsection{Dirac field}

As in the previous sections, in order to find the contribution of the
conical singularities to $\mbox{Tr}K^{(1/2)}$ for the Dirac field
$\psi$ we begin with the simple spaces. We consider the cone ${\cal
C}_\beta$ 
\begin{equation}\label{cone}
ds^2=r^2d\tau^2+dr^2~~~,~~~0\leq\tau\leq\beta~~~,
\end{equation}
where the trace of the heat kernel operator can be found in the same
way as for the spin-0 kernel, see Ref.s \cite{CKV,F94a}. As a check, in the Appendix we find the spectrum of the Dirac operator on $S^2_\beta$, and prove that the computation of
$A_1$-coefficient on this space in terms of the $\zeta$-function is in
agreement with the results obtained on ${\cal C}_\beta$. 

It is convenient to choose the following representation for the  
$\gamma$ matrices
\begin{equation}\label{D1}
\gamma_\tau=\sigma_1~~~,~~~\gamma_r=\sigma_2~~~,~~~\left\{\gamma_i,   
\gamma_j\right\}=2\delta_{ij}~~~,
\end{equation}
where $\sigma_k$ are the Pauli matrices. The covariant derivative  
is defined as
\begin{equation}\label{D2}
\nabla_{\mu}\psi=(\partial_{\mu}+\frac i2 \sigma_3~w_\mu)\psi~~~.
\end{equation}
The connection $w_\mu$ is calculated by using the tetrades which are
parallel to the polar coordinates (\ref{cone}). This gives $w=w_\mu
dx^\mu=-d\tau$. According to our general definition (\ref{bcond}), the
corresponding spinors $\psi$ on ${\cal C}_\beta$ obey the antiperiodic
conditions 
\begin{equation}\label{D3}
\psi(r,\tau+\beta)=-\psi(r,\tau)~~~.
\end{equation}
To simplify the calculation one can get rid of the connection $w_\mu$  
by the gauge-like transformation
\begin{equation}\label{D4}
\nabla_\mu\psi=\nabla_\mu\left( e^{-\frac   
i2\sigma_3\tau}\psi'\right)=e^{-\frac  i2\sigma_3\tau}\partial_\mu  
\psi'~~~,
\end{equation}
with the corresponding change of the periodicity condition (\ref{D3})  
to 
\begin{equation}\label{D5}
\psi'(r,\tau+\beta)=-e^{\frac  i2\sigma_3\beta}\psi'(r,\tau)~~~.
\end{equation}
The operator $\bigtriangleup^{(1/2)}$ acts on the transformed spinors
$\psi'$ as operator $\bigtriangleup^{(0)}$. Thus, on ${\cal C}_\beta$ one
can write the following relation 
\begin{equation}\label{D6}
\mbox{Tr}K^{(1/2)}=\mbox{Tr}K^{(0)}_{\delta_+}+
\mbox{Tr}K^{(0)}_{\delta_-}~~~,
\end{equation}
where $K^{(0)}_{\delta_{\pm}}$ are the heat kernels for the scalar  
Laplacians with the "twisted" conditions
$$
\phi(r,\tau+\beta)=-e^{\pm i \beta/2}\phi(r,\tau)\equiv
e^{i\delta_{\pm}}\phi(r,\tau)~~~,
$$ 
imposed on the fields. The form of $K^{(0)}_{\delta_{\pm}}$ was
already studied in the literature. As was shown by Dowker \cite{Dowker:94PR}, the heat kernel
$K^{(0)}_{\delta}(r,r',\tau-\tau',s)$ for the Laplacian with the more
general condition $\phi(r,\tau+\beta)=e^{i\delta}\phi(r,\tau)$  on
${\cal C}_\beta$ can be expressed in terms of the heat kernel
$K^{(0)}(r,r',\tau-\tau',s)$ on the plane $R^2$ with $\delta=0$ 
\begin{equation}
\left. K^{(0)}_{\delta}(r,r',\tau-\tau',s)\right|_{{\cal  
C}_\beta}=K^{(0)}(r,r',\tau-\tau',s)
+{1 \over 
2i\beta}\int_{A}{\exp i{(\delta-\pi) \over \beta}(\tau-\tau'+z) \over  
\sin{\pi \over \beta}(\tau-\tau'+z)}  
K^{(0)}(r,r',z,s)dz~~~.\label{D7}
\end{equation}
Note that this equation holds when $0<\delta\leq 2\pi$. The contour
$A$ lies in the complex plane and consists of two curves, going from
$-\pi +i\infty$ to $-\pi-i\infty$ and from $\pi-i\infty$ to
$\pi+i\infty$. From equation (\ref{D7}) one gets for the trace 
\begin{equation}\label{D8}
\left.\mbox{Tr}K^{(0)}_{\delta}(s)\right|_{{\cal C}_\beta}=
{\beta \over 2\pi}\left.\mbox{Tr}K^{(0)}(s)\right|_{R^2}+
{1 \over 8\pi is}\int_{0}^{\infty} rdr\int_{A}dz{\exp i{(\delta-\pi)  
\over \beta}z \over \sin
{\pi \over \beta}z}\exp{r^2\sin^2{z \over 2} \over s}~~~,
\end{equation}
which gives when one integrates first over $r$ and then over $z$
\begin{equation}\label{D9}
\left.\mbox{Tr}K^{(0)}_{\delta}(s)\right|_{{\cal C}_\beta}={\beta  
\over 2\pi}\left.\mbox{Tr}K^{(0)}(s)\right|_{R^2}+
{\beta \over 24\pi}\left(\left ({2\pi \over \beta}\right)^2 -  
1\right)-{\delta \over 4\pi\beta}(2\pi-\delta)~~~.
\end{equation}
In our case one can choose the following phase factors
\begin{equation}\label{D10}
\delta_{\pm}=\pi\pm{\beta \over 2}~~~,
\end{equation}
which is possible when $\beta\leq 2\pi$. (One can go to others values  
$\beta > 2\pi$ by means of an analytical continuation.) Then Eqs.  
(\ref{D6}), (\ref{D7}) result in the relation on ${\cal C}_\beta$
\begin{equation}\label{D11}
\left.\mbox{Tr}K^{(1/2)}(s)\right|_{{\cal C}_\beta}={\beta \over  
2\pi}\left.\mbox{Tr}K^{(1/2)}(s)\right|_{R^2}
-{\beta \over 24\pi} \left(\left ({2\pi \over \beta}\right)^2 - 1\right)~~~,
\end{equation}
which agrees with the result of Kabat \cite{Kabat}. Eq.(\ref{D11})
has the trivial consequence
\begin{equation}\label{D12}
\left. A_1^{(1/2)}\right|_{{\cal C}_\beta}=-{\beta \over 6}
\left(\left ({2\pi \over \beta}\right)^2 - 1\right)~~~.
\end{equation}
Comparing (\ref{D12}) with (\ref{i6}) one can see that the heat
coefficient $A_1^{(1/2)}$ of the spin $1/2$ Laplacian on a cone is
just the minus of the same coefficient of the scalar operator. This result for the Dirac fields can be generalized on
arbitrary manifolds with conical singularities. As before, we need to
calculate the $A_1$-coefficient on the space product ${\cal
C}_\beta\times \Sigma$ where the heat kernel operator is 
\begin{equation}
\label{D16}\left.\mbox{Tr}  
K^{(1/2)}\right|_{{\cal C}_\beta\times\Sigma}=
\left. \mbox{Tr} K^{(1/2)}\right|_{{\cal C}_\beta}
\left. \mbox{Tr} K^{(1/2)}\right|_{\Sigma}~~~.
\end{equation}
Then, by observing that $N^{(1/2)}(d-2)=\frac 12 N^{(1/2)}(d)$ and
using Eq. (\ref{D12}) we find the correction to the heat coefficient
from the conical singularities 
\begin{equation}\label{D17}
A_{\beta,1}^{(1/2)}=
-{N^{(1/2)}(d) \over 2}A_{\beta,1}^{(0)}~~~.
\end{equation}
One immediate consequence of this formula is that the known relation
between complete scalar and spinor coefficients 
\begin{equation}\label{D20}
\bar{A}_{1}^{(1/2)}=-{N^{(1/2)}(d) \over 2}~\bar{A}_{1}^{(0)}~~~,
\end{equation}
holds as well on manifolds with conical singularities. 

\subsection{Rarita-Schwinger field}
The Rarita-Schwinger field $\psi_\mu$ plays an important role in 
supergravity where it appears as gravitino, a superpartner of
graviton. If the background metric obeys the vacuum Einstein equations
the Lagrangian of Rarita and Schwinger \cite{RS} is invariant under
gauge transformations\footnote{For non-zero cosmological constant the
Rarita-Schwinger action must include an additional term, see Ref.
\cite{T}.}.  In the harmonic gauge $\gamma^\mu\psi_\mu=0$ the wave
operator for $\psi_\mu$ is reduced to $\triangle^{(3/2)}$ \cite{Das},
Eq.(\ref{i4}), and this is the reason why the latter was chosen for our
consideration. Further, we will use the following relations 
\begin{equation}\label{r1a}
\triangle^{(3/2)}(\gamma^\mu\psi)=\gamma^\mu\left((\triangle^{(1/2)}-
\Lambda)\psi\right)~~~,
\end{equation}
\begin{equation}\label{r1b}
\triangle^{(3/2)}(\nabla^\mu\psi)=\nabla^\mu\left((\triangle^{(1/2)}-
\Lambda)\psi\right)~~~,
\end{equation}
\begin{equation}\label{r1c}
\gamma^\mu\triangle^{(3/2)}\psi_\mu=(\triangle^{(1/2)}-\Lambda
)\gamma^\mu\psi_\mu~~~,
\end{equation}
\begin{equation}\label{r1d}
\nabla^\mu\triangle^{(3/2)}\psi_\mu=(\triangle^{(1/2)}-\Lambda
)\nabla^\mu\psi_\mu~~~,
\end{equation}
which hold on Einstein spaces $R_{\mu\nu}=\Lambda g_{\mu\nu}$. Then,
as in the case of the Lichnerowicz operator, we analyze the properties of
$\triangle^{(3/2)}$ on the spherical domain $S^2_\beta$. In this case it
is convenient to introduce the modified derivatives
$D_\mu=\nabla_\mu+\frac i2\gamma_\mu$ for fields with half odd-integer
spins. When acting on a spinor on $S^2_\beta$ these derivatives
commute 
\begin{equation}\label{r2}
[D_\mu,D_\nu]\psi=0~~~.
\end{equation}
We will use this fact to write for the Rarita-Schwinger field on  
$S^2_\beta$ the decomposition
\begin{equation}\label{r3}
\psi_\mu=\psi_\mu^L+\psi_\mu^T+\psi^H_\mu~~~,
\end{equation}
which is analogous to the Hodge-deRham decomposition (\ref{HDdec})  
for the vector field. Here 
\begin{equation}\label{r4}
\psi_\mu^L=D_\mu\psi~~~,~~~\psi_\mu^T=\epsilon_{\mu\nu}D^\nu\xi~~~,
\end{equation}
\begin{equation}\label{r5}
D^\mu\psi^H_\mu=0~~~,~~~\epsilon^{\mu\nu}D_\mu\psi^H_\nu=0~~~,
\end{equation}
and $\psi$ and $\xi$ are the Dirac spinors. The fields
$\psi_\mu^L$, $\psi_\mu^T$ and $\psi^H_\mu$ are orthogonal with
respect to the scalar product 
\begin{equation}\label{r6}
(\psi_1,\psi_2)=\int_{S^2_\beta}(\psi_1)^{+}_\mu(\psi_2)^\mu~~~.
\end{equation}
The orthogonality of $\psi_\mu^L$ and $\psi_\mu^T$ is the consequence  
of their definitions with the help of $D_\mu$
\begin{equation}\label{r4b}
D^\mu\psi^T_\mu=0~~~,~~~\epsilon^{\mu\nu}D_\mu\psi^L_\nu=0~~~.
\end{equation}
Now Eqs. (\ref{r1a})-(\ref{r1d}), where $\Lambda=1$, can be rewritten  
in the form
\begin{equation}\label{r7}
\triangle^{(3/2)}(D_\mu \psi)=D_\mu\left((\triangle^{(1/2)}-  
1)\psi\right)~~~,~~~D^\mu\triangle^{(3/2)}\psi_\mu=(\triangle^{(1/2)}- 
1)D^\mu\psi_\mu~~~,
\end{equation}
which enables one to relate on $S^2_\beta$ the operators
$\triangle^{(3/2)}$ and $(\triangle^{(1/2)}- 1)$. One can show that
normalizable harmonic modes $\psi_\mu^H$ on
$S^2_\beta$ are absent for any $\beta$. Therefore the trace of $K^{(3/2)}$ can be represented as 
\begin{equation}\label{r8}
\mbox{Tr} K^{(3/2)}=2 \mbox{Tr} K^{(1/2)}e^{s}-2n_k~~~,
\end{equation}
where $n_{k}\leq 2$ is the number of the so-called Killing spinors $\epsilon_i$  
which are the antiperiodic solutions of the equations
\begin{equation}\label{r9}
D_\mu\epsilon_i=0~~~.
\end{equation}
The spinors $\epsilon_i$ are also the zero-modes of the operator $(\triangle^{(1/2)}-1)$.
As follows from (\ref{r4}) there are no modes $\psi_\mu^L$ and
$\psi_\mu^T$ corresponding to $\epsilon_i$, and so they were subtracted in the r.h.s. of Eq. (\ref{r8}). As it is
shown in the Appendix, Eq. (\ref{r9}) has two independent solutions on
$S^2$ and no solutions on $S^2_\beta$ ($\beta\neq 2\pi k$). Hence, the cases $\beta=2\pi$
and $\beta\neq 2\pi$ must be considered separately. 

The situation reminds the difference of the Killing fields on $S^2_\beta$  
and on $S^2$ which was crucial for the analysis 
of the Lichnerowicz operator. This fact has a simple explanation,  
because vectors $\epsilon_i^{+}\gamma_\mu \epsilon_j$ constructed  
from $\epsilon_i$ obey automatically the Killing equation. Thus if  
some of the Killing generators are broken then there is a limitation on  
the number of spinors $\epsilon_i$. The    
relation of the Killing spinors $\epsilon_i$ and the Killing vectors $V^l$ on $S^2$ reads
\begin{equation}\label{r9b}
(V^0)_\mu=\epsilon_1^{+}\gamma_\mu  
\epsilon_1=-\epsilon_2^{+}\gamma_\mu  
\epsilon_2~~~,~~~(V^{+})_\mu=((V^{-})_\mu)^{*}=\epsilon_2^{+}\gamma_\mu 
\epsilon_1~~~,
\end{equation}
where $\epsilon_i$ are given in the Appendix, Eq. (\ref{eq:a7}),  
and $V^l$ are defined in terms of spherical harmonic $\phi^l$,
Eq. (\ref{L13}). On $S^2_\beta$ ($\beta\neq 2\pi k$) two  
generators corresponding to $V^{\pm}$ are broken and it prohibits  
solutions of (\ref{r9}) on this space. 

For this reason, the operator $\triangle^{(3/2)}$ at $\beta\neq 2\pi k$
has 4 additional non-trivial modes. Then one can follow the same line
of arguments as in the Section 3.2 and show that these modes are normalizable
and they add a finite term to $\mbox{Tr} K^{(3/2)}$ which
does not vanish at $\beta=2\pi$. On the other hand, in the each point of 
$S^2_\beta$, except the poles, the diagonal part of $K^{(3/2)}$
vanishes when $\beta\rightarrow 2\pi$. So the contribution of the additional modes in $K^{(3/2)}$
converges to a $\delta$-function on the poles and does not affect the
local quantities. 

The expression for the Schwinger-DeWitt coefficients can be found from
(\ref{r8}). In particular, one has 
\begin{equation}\label{r10}
\bar{A}_1^{(3/2)}=2\bar{A}_1^{(1/2)}+4 \cdot2\beta-4\pi\cdot 2n_k~~~.
\end{equation}
On $S^2$ $n_k=2$ and Eq.(\ref{r10}) simplifies to the equality
$A_1^{(3/2)}=2A_1^{(1/2)}$ which is in complete agreement
with the general formula (\ref{i7}). If $\beta\neq 2\pi$ the two singular
points give the correction to the heat coefficient 
\begin{equation}\label{r11}
2\left.A_{\beta,1}^{(3/2)}\right|_{{\cal C}_\beta}=\bar{A}_1^{(3/2)}-{\beta \over  
2\pi}\left.A_1^{(3/2)}\right|_{S^2}~~~,
\end{equation}
following from formula (\ref{i15}). Thus one gets
\begin{equation}\label{quation}\label{r12}
\left.A_{\beta,1}^{(3/2)}\right|_{{\cal C}_\beta}=
2\left.A_{\beta,1}^{(1/2)}\right|_{{\cal C}_\beta}
+4\beta~~~,
\end{equation}
where the quantity $\left.A_{\beta,1}^{(1/2)}\right|_{{\cal C}_\beta}$ is given by (\ref{D12}). To
generalize this equation on the space ${\cal
C}_\beta\times\Sigma$ one must decompose the field $\psi_\mu$ onto
parts normal and tangent to the surface $\Sigma$. According to this
decomposition one has 
\begin{equation}\label{R12}
\left. \mbox{Tr} K^{(3/2)}\right|_{{\cal C}_\beta\times\Sigma}=
\left. \mbox{Tr} K^{(3/2)}\right|_{{\cal C}_\beta}
\left. \mbox{Tr} K^{(1/2)}\right|_{\Sigma}
+\mbox{Tr}\left. K^{(1/2)}\right|_{{\cal C}_\beta}
\left. \mbox{Tr} K^{(3/2)}\right|_{\Sigma}~~~.
\end{equation}
Hence, the correction  
to the $A_1$-coefficient due to the conical singularities reads 
\begin{eqnarray}  
A_{\beta,1}^{(3/2)}&=&\left[N^{(1/2)}(d-2)\left.  
A_1^{(3/2)}\right|_{{\cal C}_\beta}+N^{(3/2)}(d-2)\left.  
A_1^{(1/2)}\right|_{{\cal C}_\beta}\right]\int_{\Sigma}
\nonumber\\
&=&-\frac 12 N^{(3/2)}(d)A_{\beta,1}^{(0)}+2\beta  
N^{(1/2)}(d)\int_{\Sigma}~~~,\label{R15}
\end{eqnarray}
which is the result reported in (\ref{i11}). Note that in Eq. (\ref{R15}) 
the simple relation 
\begin{equation}\label{R16}
2N^{(1/2)}(d-2)+N^{(3/2)}(d-2)=\frac 12 N^{(3/2)}(d)
\end{equation}
was used.

\section{Discussion}
\setcounter{equation}0

\subsection{Comparison with "blunt" cones}

In some physical problems conical singularities appear only as an 
idealization of the properties of smooth manifolds.
It reflects the simple fact that one can describe a singular space  
${\cal M}_\beta$ as a convergent sequence of manifolds ${\tilde {\cal  
M}}_\beta$ with the "blunted" conical singularities. This is the way, for  
instance, how one can define the integral geometrical characteristics  
of ${\cal M}_\beta$ constructed from the powers of the Riemann tensor  
\cite{FS95b}. In particular, this procedure gives the following  
well-known result for the integral curvature $\int {\tilde R}$ of  
${\cal M}_\beta$
\begin{equation}\label{dis1}
\int_{{\cal M}_\beta} {\tilde R}\equiv \lim_{{\tilde {\cal  
M}}_\beta\rightarrow {\cal M}_\beta}\int_{{\tilde {\cal  
M}}_\beta}R=\int_{{\cal M}_\beta-\Sigma} R+2(2\pi-\beta)\int_{\Sigma}~~~,
\end{equation}
where $R$ is the standard scalar curvature calculated on the smooth  
domain ${\cal M}_\beta-\Sigma$. 

On the other hand, the general form of the first Schwinger-DeWitt  
coefficient in the asymptotic expansion on the smooth  
manifolds is defined by the integral curvature\footnote{To be more precise it is true on manifolds without  
boundaries.}
\begin{equation}\label{dis2}
A_1^{(j)}=c^{(j)}\int R~~~,
\end{equation}
where the coefficients $c^{(j)}$ depend on the spin $j$ and can be found  
from Eq.(\ref{i7}): 
\begin{equation}\label{dis5}
c^{(0)}=\frac 16~,~c^{(1/2)}=-{N^{(1/2)}\over 12}~,~
c^{(1)}={N^{(1)}\over 6}-1~,~c^{(3/2)}=-{N^{(3/2)}\over 12}~,~
c^{(2)}={N^{(2)}\over 6}-(d+2)~.
\end{equation}
Therefore if the integral curvature is calculated  
as the limit (\ref{dis1}),
the coefficients (\ref{dis2})  have the finite values
$\tilde{A}_1^{(j)}$ on the singular space ${\cal M}_\beta$  
\begin{equation}\label{dis3}
\tilde{A}_1^{(j)}=\lim_{{\tilde {\cal M}}_\beta\rightarrow {\cal  
M}_\beta}A_1^{(j)}[{\tilde {\cal M}}_\beta]=A_1^{(j)}+c^{(j)}
2(2\pi-\beta)\int_{\Sigma}~~~,
\end{equation}
where $A_1^{(j)}$ is given by the integral (\ref{dis2}) over the
domain ${\cal M}_\beta-\Sigma$. It is worth comparing $A_1$-coefficients (\ref{dis3})
computed on the manifolds ${\tilde {\cal  
M}}_\beta$ with the blunted singularities and the results
(\ref{i8})-(\ref{i12}) obtained by the direct computation of ${\mbox
Tr}K^{(j)}$ on ${\cal M}_\beta$. In the both cases the coefficients have the
similar structures and the conical singularities add the
surface terms. However, the contributions $A_{\beta,1}^{(j)}$ given by
(\ref{i9})-(\ref{i12}) and the surface terms in (\ref{dis3}) are
different.  Only for spins $j=0$, $1/2$ and $1$ and only in the limit
of small deficits of the conical angle one has a correspondence
\begin{equation}\label{dis4}
A_{\beta,1}^{(j)}=c^{(j)}  
2(2\pi-\beta)\int_{\Sigma}+O\left((2\pi-\beta)^2\right)~~~,~~~j=0,~
\frac 12,~1~~~,
\end{equation}
which holds up to the terms of the second order in $(2\pi-\beta)$.  
Thus, as it follows from (\ref{i8}) and (\ref{dis3}), the relation  
between the complete coefficients reads
\begin{equation}\label{dis6}
\bar{A}_1^{(j)}=\tilde{A}_1^{(j)}+O\left((2\pi-\beta)^2\right)~~~,~~~j 
=0,~\frac 12,~1~~~.
\end{equation}
On the contrary, 
for spins 3/2 and 2 the relation (\ref{dis6}) does not hold, because the surface corrections in these cases do not vanish at
$\beta=2\pi$. This disagreement occurs because the
local isometries of the blunted manifolds ${\tilde {\cal  
M}}_\beta$ are not broken by the
singularities. Therefore the heat kernels for spins 3/2 and 
2 on ${\tilde {\cal  M}}_\beta$ cannot be used as the approximation of the corresponding kernels on the singular manifolds ${\cal M}_\beta$ even when ${\tilde {\cal  M}}_\beta\rightarrow {\cal M}_\beta$.

\subsection{One-loop ultraviolet divergencies}

Let us
consider now the one-loop effective action $W^{(j)}$ for a spin 
$j$ on a curved background. In the Schwinger-DeWitt representation it  
looks as
\begin{equation}\label{dis7}
W^{(j)}=(-1)^{2j}\frac 12 \log\det \triangle^{(j)}=
-(-1)^{2j}\frac 12\int^{\infty}_{\delta^2}{ds \over s} {\mbox Tr}
K^{(j)}(s)~~~,
\end{equation}
where $\delta^2$ stands for an ultraviolet cut-off and the factor $(-1)^{2j}$ is related to the statistics. The structure of the  
ultraviolet divergences of $W^{(j)}$ is determined by the asymptotic  
behavior of ${\mbox Tr}K^{(j)}(s)$ at small $s$, where
one can use the asymptotic expansion (\ref{i1}). In particular, the  
divergence $W^{(j)}_{\mbox{div},1}$ related to the first heat  
coefficient $A_1^{(j)}$ is
\begin{equation}\label{dis8}
W^{(j)}_{\mbox{div},1}=-{(-1)^{2j} \over 32\pi ^2\delta  
^2}A_1^{(j)}~~~.
\end{equation}
According to Eq.(\ref{dis2}), on the smooth manifolds $A_1^{(j)}$ is  
proportional to the integral of the scalar curvature $R$, and so the  
divergence (\ref{dis8}) is removed by the renormalization of 
the Newton constant $G$ in the bare gravitational action
${1 \over 16\pi G} \int R$, see for instance \cite{BirDev}.

In the last years much attention has been paid to the same
renormalization problem on manifolds ${\cal M}_\beta$ with conical
singularities \cite{SU}-\cite{LW}. To discuss this problem we will
follow the line of arguments of \cite{FS95a}. It is reasonable to
assume that the bare gravitational action is determined by the total integral
curvature (\ref{dis1}) of ${\cal M}_\beta$, which is the limiting value
of the curvature on the blunted spaces $\tilde{{\cal M}}_\beta$. Thus,
by taking into account Eqs.(\ref{dis6}) and (\ref{dis8}), one can write for spins
$j=0,~1/2,~1$ on singular spaces the following chain of relations 
\begin{eqnarray}
{1 \over 16\pi G_{\mbox{bare}}}\int_{{\cal M}_\beta}  
\tilde{R}+W^{(j)}_{\mbox{div},1}[{\cal M}_\beta]
=\lim_{{\tilde {\cal M}}_\beta\rightarrow {\cal M}_\beta}\left(
{1 \over 16\pi G_{\mbox{bare}}}\int_{\tilde{{\cal M}}_\beta}R+
W^{(j)}_{\mbox{div},1}[\tilde{{\cal M}}_\beta]\right)
\nonumber\\
+ O\left((2\pi-\beta)^2\right)
={1 \over 16\pi G_{\mbox{ren}}}\int_{{\cal M}_\beta} \tilde{R}+
O\left((2\pi-\beta)^2\right)~~~,~~~j=0,~\frac 12,~1~~~.
\label{dis9}
\end{eqnarray}
The connection between the bare $G_{\mbox{bare}}$ and renormalized
$G_{\mbox{ren}}$ constants is standard because $\tilde{{\cal
M}}_\beta$ are smooth manifolds. It means that for spins $j=0,~1/2,~1$
the standard renormalization of the gravitational constant removes the
divergences up to the terms of the second order in $(2\pi-\beta)$.
This property, however, is not true for spins $j=3/2$ and $2$. 

\subsection{Off-shell calculations of the entropy on black-hole  
backgrounds}

We now briefly discuss our results in connection with off-shell
calculations of the entropy on black hole instantons with conical
singularities. The off-shell methods are required for the
statistical-mechanical computations in quantum theory on black-hole
backgrounds (a review of off-shell approaches can be found in
\cite{FFZ}). In the Euclidean formulation of the gravitational
thermodynamics \cite{GiHa:76},\cite{Hawk:79} the fields are taken on
the Euclidean section of the corresponding Lorentzian manifold. The
imaginary time period $\beta$ is associated with the inverse
temperature. In the case of black holes the Euclidean instanton 
${\cal M}_\beta$ has the
conical singularities if $\beta\neq 2\pi$. The regularity condition
$\beta=2\pi$ at the Euclidean horizon is, at the same time, the
condition of the thermal equilibrium of the black hole and its radiation. 

The free-energy for a spin $j$ is proportional to the one-loop
effective action $W^{(j)}$, Eq.(\ref{dis7}), and the contribution
$S^{(j)}$ of the given field into the entropy is 
\begin{equation}\label{dis10}
S^{(j)}=\left. \left(\beta{\partial \over \partial  
\beta}-1\right)W^{(j)}\right|_{\beta=2\pi}~~~,
\end{equation}
where the derivative is taken over the period of the singular
instanton ${\cal M}_\beta$. The divergent part
$W^{(j)}_{\mbox{div},1}$ of the action $W^{(j)}$ on ${\cal M}_\beta$ results in the
divergent  correction $S^{(j)}_{\mbox{div},1}$ to $S^{(j)}$
proportional to the horizon area $\int_\Sigma$: 
\begin{equation}\label{dis10a}
S^{(j)}_{\mbox{div},1}=(-1)^{2j}{c^{(j)} \over 8\delta^2}\int_\Sigma~~,~~\mbox{for}~~j=0,~1/2,~1,~3/2~~~,
\end{equation}
\begin{equation}\label{dis10b}
S^{(2)}_{\mbox{div},1}={1 \over 8\delta^2}(c^{(2)}+2)\int_\Sigma~~~,
\end{equation}
where $c^{(j)}$ are defined in (\ref{dis5}).
Formally this effect
occurs because of the conical singularities, and it drawn a
considerable interest in the literature \cite{SU}-\cite{LW} because
$S^{(j)}_{\mbox{div},1}$ has the same form as the mysterious
Bekenstein-Hawking entropy $S_{BH}={1 \over 4G}\int_{\Sigma}$ \cite{Beke},\cite{Hawk:75}. This has the
important consequence. As follows from (\ref{dis9}) the surface
divergences $W^{(j)}_{\mbox{div},1}$ removed under standard
renormalization of the gravitational constant up to terms
$O\left((2\pi-\beta)^2\right)$ which do not contribute to the entropy
(\ref{dis10}). Thus, for spins $j=0,~1/2,~1$ the correction
$S^{(j)}_{\mbox{div},1}$  renormalizes the Bekenstein-Hawking entropy $S_{BH}={1 \over 4G}\int_{\Sigma}$, see \cite{SU}-\cite{LW}.  

For spin 3/2 the entropy divergences can be also renormalized because the part of $A_{\beta,1}^{(3/2)}$, which does not vanish at $\beta=2\pi$, is proportional
to $\beta$ (see Eq. (\ref{i11})), and so it does not contribute to the entropy calculated by formula (\ref{dis10}). However for this spin
the non-renormalizable correction appear in the energy $E^{(j)}={\partial \over\partial
\beta} W^{(j)}$.
Finally, our analysis shows that the renormalization of the
off-shell one-loop corrections to the entropy does not work for tensor
fields.  The origin of this result is in
the specific properties of the Lichnerowicz 
operator. 

The complete investigation of the
renormalization problem for the graviton and gravitino  must also
take into account the ghosts, whose wave operators are similar
to the vector $\triangle^{(1)}$ and spinor $\triangle^{(1/2)}$
Laplacians. Note, however, that in the conical singularity method the study of the quantum corrections to the black hole entropy from the gravitons may be non-trivial problem. In this case one must first formulate the quantum theory for the metric perturbations on singular backgrounds.
We will return to this question in the next section. 

\bigskip 

In the end a small comment about entropy of the Maxwell field on
two-dimensional Rindler-like spaces is in order. This question was
discussed in \cite{Kabat}. As it was pointed out in this paper, there
are no dynamical degrees of freedom in two dimensional Maxwell theory
because of two gauge constraints. For this reason any kind of entropy 
for the Maxwell field must vanish. 

The results of Section 3.1 can be used to show that the off-shell
calculation of the entropy on two-dimensional manifolds with conical
singularities is in agreement with that general observation. Indeed,
in the Feynmann gauge $\nabla^\mu V_\mu=0$ the one-loop effective action $W_{\mbox{gauge}}$ of the abelian gauge field is determined
as 
\begin{equation}\label{dis11}
\exp(-W_{\mbox{gauge}})={\det '(\triangle^{(0)})~ \over 
\left(\det '(\triangle^{(1)})\right)^{1/2}}~~~.
\end{equation}
Eq. (\ref{dis11}) can be obtained from the corresponding functional
integral which includes the Faddeev-Popov ghosts. For the
considered gauge the
wave operator for the ghosts is the scalar Laplacian $\triangle^{(0)}$.
The determinants in
(\ref{dis11}) with the prime appear from the Gauss integrals over the vector and ghost fields and 
do not include the zero modes modes. Consequently, the
Schwinger-DeWitt representation (\ref{dis7}) for $W_{\mbox{gauge}}$
reads 
\begin{equation}\label{dis12}
W_{\mbox{gauge}}=
-\frac 12\int^{\infty}_{\delta}{ds \over s} \left[{\mbox Tr}
K^{(1)}(s)-n_1-2({\mbox Tr}K^{(0)}(s)-n_0)\right]~~~,
\end{equation}
where $n_0$ and $n_1$ is the number of vector and scalar zero modes.
In two dimensions, however, the traces of the spin 1 and spin 0
Laplacians are related by Eq.(\ref{12}) and the effective action for the
gauge field vanish: $W_{\mbox{gauge}}=0$. This result does not depend
on the background metric and also holds on manifolds with conical
singularities. Therefore, the off-shell entropy obtained from
$W_{\mbox{gauge}}$ with the help of (\ref{dis10}) is zero for 
two-dimensional abelian gauge field. This result agrees with the general
requirement. In our mind, the different conclusion that has been made
in \cite{Kabat} does not take properly into account the zero modes. 

\section{Conclusions}
\setcounter{equation}0

In this paper we study the heat kernels of the Laplace operators
which appear under quantization of non-zero spin fields on manifolds with
conical singularities. The two-dimensional domains are the
simplest arena where one can obtain the heat kernels explicitly and
understand some of their general features. Our
main conclusion is that the properties of the operators for spins 1/2
and 1 are very similar to the properties of the scalar Laplacian
considered in the literature earlier.  However, studying the spins 3/2
and 2 brings something new. The eigen-functions of the wave operators
for these spins are sensitive to the isometries of the
background space. This can be very well illustrated by examining these
operators on the simplest spherical domains $S^2_\beta$. There the
spin 2 eigen-modes constructed with the help of the Killing vectors or
spin 3/2 modes obtained from the Killing spinors are identically zero on $S^2$ and the corresponding eigen-values do not appear
in the spectrum. Conical singularities break the isometries of $S^2$
and introduce new modes. Interestingly, the contribution of these
modes into the trace of the heat kernel operator doesn't vanish even
in the limit when the conical deficit tends to zero. This happens,
however,  only on the singular points, but outside them, no matter how
close, the discrepancy with the heat kernels on the smooth spaces is
absent. This picture is also true for arbitrary singular spaces with
the structure ${\cal C}_\beta\times \Sigma$ near the
hypersurface $\Sigma$ where the conical singularities break the local
translational isometries. 

The way in which the conical singularities change the form of the
first Schwinger-DeWitt coefficient in the asymptotic expansion of the
heat kernel operator has been determined. We carried out the analysis
on $S^2_\beta$ and then generalized it to higher dimensions. As a
check of these results, it would be useful to investigate explicitly
the heat kernel expansion for spin 3/2 and 2 operators on singular
spaces with dimension higher than 2 and confirm our results. This is a subject for further analysis. 

One of the applications of our results is the quantization of 
gravity (and supergravity) in the presence of conical singularities.
This problem is beyond the scope of the present paper, but some
remarks are in order. Taking into account the properties of the heat
kernels $K^{(2)}$ and $K^{(3/2)}$ one can expect that the effective
action of the graviton and gravitino on spaces with conical
singularities will not be  reduced to the action on the regular
manifolds. The ghosts which appear under the quantization are
described by the vector and spinor fields and do not seem to change
this conclusion. Therefore, one can speculate that quantization of the
metric perturbations on singular and smooths backgrounds may be quite
different and may not coincide to each other even in the limit
$\beta\rightarrow 2\pi$. 

In connection with this problem it is worth pointing out the canonical  
formulation of gravity in the presence of conical defects which was suggested recently by Carlip and Teitelboim \cite{CaTe},\cite{Te} and used for the explanation of the Bekenstein-Hawking entropy. By analyzing the action principle, the authors showed that the deficit angle and
the area of $\Sigma$ are canonical conjugates. If these variables are quantized, then the naive limit
$\beta\rightarrow 2\pi$ in such quantum theory seems to be
inconsistent. From this point of view the disagreement between the determinants for spin 2 field on singular and smooth backgrounds would not lead to a contradiction. 
This problem is an interesting subject for further
research. 

\bigskip\bigskip
{\bf Acknowledgements}:\ \ D.V.F. is very grateful to Valeri Frolov  
and Andrei Zelnikov for helpful discussions, and G.M. would like to thank
Giovanni Sparano, Gianpiero Mangano and Giampiero Esposito for valuable 
comments and suggestions.
This work was supported in part by the Natural Sciences and  
Engineering Research Council of Canada.
\newpage
\appendix

\section{Dirac field on  $S^2_\beta$}
\setcounter{equation}0 
In this Appendix, in order to check the validity of Eq. (\ref{i9})
obtained in Sec. 4.1 , we study the Dirac operator
$\gamma^\mu\nabla_\mu$ on $S^2_\beta$. On this space
$\gamma^\mu\nabla_\mu$ has the following eigen-values 
\begin{equation}\label{a1}
\pm i \lambda_{n,m} = \pm i 
\left(n +{2 \pi \over \beta}m + {\pi \over \beta} + {1 \over 2}  
\right)~~~,
\end{equation}
where $n,m=0,1,..$ and each eigen-value has double degeneracy.
This spectrum can be used to study the $\zeta$-function for theoperator  
$\triangle^{(1/2)}=-(\gamma^\mu\nabla_\mu)^2$
\begin{equation}
\zeta^{(1/2)}(z)=4\sum_{n,m=0}^{\infty} \lambda_{n,m}^{-2z}~~~.
\label{eq:a2}
\end{equation}
In particular, this enables one to find $A_1$-coefficient, using the  
formula
\begin{equation}
\bar{A}_1^{(1/2)}=4 \pi \zeta^{(1/2)}(0)~~~,
\label{eq:a2c}
\end{equation}
and check that Eq. (\ref{eq:a2c}) agrees with Eq.(\ref{D20}) obtained
by using the heat kernel on ${\cal C}_\beta$. To this aim we will
follow approach of \cite{a2} and study first the Killing spinors which
are the solutions of Eq. (\ref{r9}) 
\begin{equation}
D_\mu\epsilon_{j}=\left(\nabla_\mu+ \frac i2  
\gamma_\mu\right)\epsilon_{j}=0~~~,
\label{eq:a3}
\end{equation}
where $\nabla_{\mu} \psi = \partial_{\mu} \psi + \frac i2 \sigma_3
\omega_{\mu} \psi$ and  $\omega=-\sin{\theta}~d\tau$. It is easy to see
that in general Eq. (\ref{eq:a3}) admits two independent solutions 
\begin{equation}
\epsilon_1(\tau,\theta) =e^{i \tau/  2}
\left[ \begin{array} {c}
\sin(\theta/2+\pi/4)
\\ \\
-\cos(\theta/2+\pi/4)
\end{array}
\right]~~~,~~~
\epsilon_2(\tau,\theta) =e^{-i \tau/  2}
\left[ \begin{array} {c}
\cos(\theta/2+\pi/4)
\\ \\
\sin(\theta/2+\pi/4)
\end{array}\right]~~~.
\label{eq:a7}
\end{equation}
These spinors obey the following conditions
\begin{equation}\label{a7}
\epsilon_{1}(\tau+\beta,\theta)=e^{i\beta / 2}
\epsilon_{1}(\tau,\theta)~~~,~~~
\epsilon_{2}(\tau+\beta,\theta)=
e^{-i\beta / 2} \epsilon_{2}(\tau,\theta)~~~,
\end{equation}
and are normalized as $\epsilon^{\dag}_i\epsilon_j=\delta_{ij}$ with
$i,j=1,2$. Thus the antiperiodic solutions of (\ref{eq:a3}) exist only
on $S^2$ (and, more generally, at $\beta=2\pi k$). According to the
method described in \cite{a2}, the eigen-functions $\psi_\lambda$ of
the Dirac operator 
\begin{equation}
\gamma^{\rho} \nabla_{\rho} \psi_\lambda =  i\lambda~
\psi_\lambda ~~~,
\label{eq:a10}
\end{equation}
can be represented as the linear combinations 
\begin{equation}\label{d10}
\psi_\lambda=[i\lambda  
\phi_\lambda+\gamma^{\mu}(\partial_\mu\phi_\lambda)]\epsilon_i~~~, 
\end{equation} 
\begin{equation}\label{d11}
\psi_{-\lambda-1}=[-i(\lambda+1)  
\phi_\lambda+\gamma^{\mu}(\partial_\mu\phi_\lambda)]\epsilon_i~~~, 
\end{equation}
constructed in terms of the spinors $\epsilon_i$ and the eigen-functions
$\phi_{\lambda}$ of the scalar operator $\Delta^{(0)}$ 
\begin{equation}
\Delta^{(0)} \phi_{\lambda} = \lambda(\lambda+1)\phi_{\lambda}~~~.
\label{eq:a8}
\end{equation}
The periodicity conditions for scalar functions $\phi_\lambda$ must be
chosen in such a way to get antiperiodic eigen-vectors for the Dirac
operator $\psi_\lambda(\tau+\beta)=-\psi_\lambda(\tau)$. Let us denote
by $\phi_{\lambda}$ and $\tilde{\phi}_{\lambda}$ the scalar modes
corresponding to $\epsilon_1$ and $\epsilon_2$ spinors,
respectively. Then they satisfy the following conditions 
\begin{equation}
\label{eq:a13a}
\phi_\lambda(\tau+\beta,\theta)= -e^{-i \beta/2}
\phi_\lambda(\tau,\theta)~~~,~~~ 
\tilde{\phi}_\lambda(\tau+\beta,\theta)= 
-e^{i \beta/2}\tilde{\phi}_\lambda(\tau,\theta)~~~. 
\end{equation}
The both modes $\phi_{\lambda}(\tau,\theta)$ and 
$\tilde{\phi}_{\lambda}(\tau,\theta)$ admit the same eigen-values.
In particular, one gets the eigen-values  
$\lambda_{n,m}(\lambda_{n,m}+1)$ with the double degeneracy, where
\begin{equation}\label{d2}
\lambda_{n,m}=n+{(2m+1)\pi \over \beta}+\frac12~~~,~~~n,m=0,1,2,...
\end{equation}
whose two corresponding eigen-functions are 
\begin{equation}
\phi_{n,m}^{(1)} =\phi_{n+1,m}~~~,~~~\phi_{n,m}^{(2)} =  
\phi_{n,-m-1}~~~,
\label{d3}
\end{equation}
for $\phi_{\lambda}(\tau,\theta)$ modes and 
\begin{equation}
\tilde{\phi}_{n,m}^{(1)}=\tilde{\phi}_{n,m}~~~,~~~
\tilde{\phi}_{n,m}^{(2)}=\tilde{\phi}_{n+1,-m-1}~~~,\label{d5a}
\end{equation}
for $\tilde{\phi}_{\lambda}(\tau,\theta)$ modes.
The functions $\phi_{n,m}$ and $\tilde{\phi}_{n,m}$ are defined as
\begin{eqnarray}
\phi_{n,m}(\tau,\theta) & \equiv & 
\exp\left( i q \tau \right)
~\left( \cos{\theta}\right)^{|q|}~
P^{(|q|,|q|)}_{n}(\sin\theta)~~~,\label{eq:a23}\\
\tilde{\phi}_{n,m}(\tau,\theta) & \equiv & 
\exp\left( i \tilde{q} \tau \right)
~\left( \cos{\theta}\right)^{|\tilde{q}|}~
P^{(|\tilde{q}|,|\tilde{q}|)}_{n}(\sin\theta)~~~, 
\label{eq:b23a}
\end{eqnarray}
where $P^{(|q|,|q|)}_n(x)$ stand for Jacobi polynomials, $n,|m|=0,1,...$, and 
\begin{equation}
\label{eq:a15}
q\equiv {(2 m+1)\pi \over \beta} - {1 \over 2}~~~,~~~
\tilde{q}\equiv{(2 m+1)\pi \over \beta} + {1 \over 2}~~~.
\end{equation}
The operator $\triangle^{(0)}$ also has 
non-degenerate eigen-values $\lambda_m(\lambda_m+1)$, where
\begin{equation}
\label{d4}
\lambda_m={(2m+1)\pi \over \beta}-\frac 12~~~,~~~m=0,1,2,...~~~,
\end{equation}
whose corresponding eigen-vectors are
\begin{equation}
\label{d5}
\phi_m=\phi_{0,m}~~~,
~~~\tilde{\phi}_m=\tilde{\phi}_{0,-m-1}~~~.
\end{equation}
For the Dirac operator on $S^2_\beta$ the eigen-values $\pm 
i\lambda_{n,m}$ can have double degeneracy, because the
corresponding eigen-value problem consists of two first order
differential equations. The eigen-vectors constructed by means of
spinors $\epsilon_1$ and $\epsilon_2$ have the same eigen-values.
Therefore for double degenerate modes corresponding to eigen-values
$\pm i\lambda_{n,m}$, Eq. (\ref{d2}), one can only use one of these
spinors, for instance $\epsilon_1$. In this case we can use the
property $\lambda_{n,m}+1=\lambda_{n+1,m}$, and represent the
corresponding modes in the form 
\begin{equation}\label{d12}
\psi_{n,m}^{(+,l)}=\left[ i\lambda_{n,m}  
\phi^{(l)}_{n,m}+\gamma^{\mu}(\partial_\mu\phi^{(l)}_{n,m})\right]
\epsilon_1~~~,~~n,m=0,1,2,..,
\end{equation}
\begin{equation}\label{d13}
\psi_{n,m}^{(-,l)}=\left[ -i\lambda_{n,m}  
\phi^{(l)}_{n-1,m}+\gamma^{\mu}(\partial_\mu  
\phi^{(l)}_{n-1,m})\right]
\epsilon_1~~~,~~n=1,2,...~~~m=0,1,2,..,
\end{equation}
where $\phi^{(l)}_{n,m}$ with $l=1,2$ denote the scalar solutions
(\ref{d3}). Functions (\ref{d12}), (\ref{d13}) obey the equations 
\begin{equation}\label{d14}
\gamma^\mu\nabla_\mu\psi_{n,m}^{(\pm,l)}=\pm  
i\lambda_{n,m}\psi_{n,m}^{(\pm,l)}~~~.
\end{equation}
Other eigen-functions can be obtained by making use of non-degenerate  
scalar modes $\phi_m$ and $\tilde{\phi}_m$ for which we have the  
identities:
\begin{equation}\label{d16a}
\left[i\lambda_m\phi_m+\gamma^{\mu}(\partial_\mu\phi_m)\right]
\epsilon_1=0~~~,~~~
\left[i\lambda_m\tilde{\phi}_m+\gamma^{\mu}(\partial_\mu
\tilde{\phi}_m)\right]\epsilon_2=0~~~,
\end{equation}
following from (\ref{eq:a23}) and (\ref{eq:b23a}). They show that
spinors $\phi_m \epsilon_1$ and
$\gamma^{\mu}(\partial_\mu\phi_m)\epsilon_1$ ( $\tilde{\phi}_m
\epsilon_2$ and $\gamma^{\mu}(\partial_\mu\tilde{\phi}_m)\epsilon_2$)
are not independent, and so the only combinations one can construct from
them are 
\begin{equation}\label{d15}
\psi_{0,m}^{(-,1)}=\phi_{m}\epsilon_1~~~,~~~
\psi_{0,m}^{(-,2)}=\tilde{\phi}_{m}\epsilon_2~~~.
\end{equation}
As the consequence of (\ref{d16a}), $\psi_{0,m}^{(-,1)}$ and
$\psi_{0,m}^{(-,2)}$ have the coinciding eigen-values $-i(\lambda_m
+1)=-i\lambda_{0,m}$. Finally one can check the orthogonality of the
modes (\ref{d12}), (\ref{d13}) and (\ref{d15}).  By gathering all
the 
eigen-values one gets the spectrum of the Dirac operator on $S^2_\beta$
in the form (\ref{a1}). At $\beta=2\pi$ it reproduces the spectrum of
this operator on $S^2$ \cite{Camporesi}. 

To calculate the zeta-function
(\ref{eq:a2}) it is suitable to write $\zeta^{(1/2)}$ in the
different form (cf \cite{FM}) 
\begin{equation}\label{eq:a24b}
\zeta^{(1/2)}(z) = 4  
\sum_{m=0}^{\infty}
\zeta_{R}\left(2z,\alpha m + \gamma \right)= { 4 \over \Gamma(2z)} 
\int_{0}^{\infty} { y^{2z-1} \over 1 - e^{-y}} e^{- \gamma y} \left(
{1 \over 1 - e^{- \alpha y} }\right) ~dy~~~,
\end{equation}
where $\gamma\equiv (\pi / \beta) + (1 / 2)$, $\alpha=2 \pi / \beta$
and $\zeta_{R}(z)$ is the Riemann zeta-function. Then one can decompose the
quantity $(1 - e^{- \alpha y})^{-1}$ in the series in powers of $\alpha y$.
After that the simple calculation gives 
\begin{equation}\label{eq:a25}
\zeta^{(1/2)}(0) = - { 4 \over \alpha} 
\zeta_R(-1,\gamma) - 4 B_1 \zeta_R(0,\gamma) + 2\alpha B_2
=-\left({\beta \over 12 \pi}+{\pi \over 3 \beta}\right)~~~,
\end{equation}
where $B_n$ are the Bernoulli numbers. As one can check now, Eq.  
(\ref{eq:a25}) gives the value of $A_1$-coefficient on $S^2_\beta$  
defined by Eq.(\ref{eq:a2}) which exactly coincides
with the results obtained in Sec. 4.1 by the different method
using the heat kernel operator on ${\cal C}_\beta$.  

\end{document}